\documentclass{ifacconf}

\usepackage{graphicx}      
\usepackage{natbib}        
\usepackage{tikz}
\usetikzlibrary{calc,shapes,arrows,automata}
\usepackage{tabularx}
\usepackage{booktabs} 
\usepackage[inkscapeformat=png]{svg}
\newcommand{\R}{\mathbb{R}}
\newcommand{\N}{\mathbb{N}}

\newcommand{\Xx}{\mathcal{X}}

\newcommand{\Sys}{\mathfrak{S}}

\newcommand{\Tt}{\mathcal{T}}

\newcommand{\hf}{\hat{f}}

\newcommand{\seq}[1]{\langle #1 \rangle}

\usepackage{placeins}
\usepackage{algorithm}
\usepackage{algpseudocode}
\usepackage{tikz}
\usetikzlibrary{calc,shapes,arrows,automata}
\usepackage{amsmath}
\usepackage{url}
\usepackage{subcaption} 
\usepackage{caption}
\usepackage[utf8]{inputenc}
\usepackage{latexsym,amsfonts}
\usepackage{mdframed}
\usepackage{caption}
\usepackage{threeparttable}
\usepackage{mathtools}

\usepackage{newfloat}
\usepackage{listings}
\DeclareOldFontCommand{\rm}{\normalfont\rmfamily}{\mathrm}
\DeclareOldFontCommand{\sf}{\normalfont\sffamily}{\mathsf}
\DeclareOldFontCommand{\tt}{\normalfont\ttfamily}{\mathtt}
\DeclareOldFontCommand{\bf}{\normalfont\bfseries}{\mathbf}
\DeclareOldFontCommand{\it}{\normalfont\itshape}{\mathit}
\DeclareOldFontCommand{\sl}{\normalfont\slshape}{\@nomath\sl}
\DeclareOldFontCommand{\sc}{\normalfont\scshape}{\@nomath\sc}
\def\qed{\relax\ifmmode\hskip2em \Box\else\unskip\nobreak\hskip1em $\Box$\fi}
\def\proof@headerfont{\upshape\bfseries}
 \gdef\th@plain{\itshape
  \def\@begintheorem##1##2{\item[\hskip\labelsep
    {\theorem@headerfont ##1\ ##2.}]}%
  \def\@opargbegintheorem##1##2##3{\item[\hskip\labelsep
    {\theorem@headerfont ##1\ ##2}\ {\upshape (##3).}]}}
\gdef\th@definition{\upshape
  \def\@begintheorem##1##2{\item[\hskip\labelsep
    {\theorem@headerfont ##1\ ##2.}]}%
  \def\@opargbegintheorem##1##2##3{\item[\hskip\labelsep
    {\theorem@headerfont ##1\ ##2}\ {\upshape(}{\it ##3\/}{\upshape).}]}}
\gdef\theorem@headerfont{\itshape}
\gdef\th@plain{\upshape
  \def\@begintheorem##1##2{\item[\hskip\labelsep
    {\theorem@headerfont ##1\ ##2.}]}%
  \def\@opargbegintheorem##1##2##3{\item[\hskip\labelsep
    {\theorem@headerfont ##1\ ##2.}\ {\upshape (##3).}]}}
\gdef\th@definition{\upshape
  \def\@begintheorem##1##2{\item[\hskip\labelsep
    {\theorem@headerfont ##1\ ##2.}]}%
  \def\@opargbegintheorem##1##2##3{\item[\hskip\labelsep
    {\theorem@headerfont ##1\ ##2.}\ {\upshape (##3).}]}}

\def\rom#1{\leavevmode\skip@\lastskip\unskip\/%
        \ifdim\skip@=\z@\else\hskip\skip@\fi
   {\upshape#1}}
\theorempostskipamount=\theorempreskipamount
\newenvironment{pf}%
  {\par\addvspace{\theorempreskipamount}\noindent
   {\proof@headerfont\Elproofname}\enspace\ignorespaces}%
  {\par\addvspace{\theorempreskipamount}}
\def\Elproofname{Proof.}
\expandafter\let\csname endpf*\endcsname=\endpf

 \theoremstyle{plain}
\if@secthm
  \newtheorem{thm}{Theorem}[section]
  \@addtoreset{thm}{section}
\else
  \newtheorem{thm}{Theorem}
\fi

\newtheorem{assum}[thm]{Assumption}

 \theoremstyle{definition}
\newtheorem{defn}[thm]{Definition}

\long\def\@makealgocaption#1#2{\vskip 2ex \small
  \hbox to \hsize{\parbox[t]{\hsize}{{\bfseries #1.} #2}}}

\def\@pnumwidth{2.55em}
\def\@tocrmarg{2.55em \@plus 5em}
\def\@dotsep{-2.5}
\usepackage[normalem]{ulem}

\begin{document}
\begin{frontmatter}

\title{Transfer of Safety Controllers Through Learning Deep Inverse Dynamics Model} 

\thanks[footnoteinfo]{This work was supported in part by the NSF through the NSF CAREER awards CCF-2146563, and CNS-2145184, and is supported by grants CNS-2039062 and CNS-2111688.}

\author{Alireza Nadali, Ashutosh Trivedi, and Majid Zamani}

\address{University of Colorado Boulder, 
   Boulder, CO 80301 USA \\(e-mail: \{a\_nadali, ashutosh.trivedi, majid.zamani\}@ colorado.edu).}

\begin{abstract}          

Control barrier certificates have proven effective in formally guaranteeing the safety of the control systems. 
However, designing a control barrier certificate is a time-consuming and computationally expensive endeavor that requires expert input in the form of domain knowledge and mathematical maturity. Additionally, when a system undergoes slight changes, the new controller and its correctness certificate need to be recomputed, incurring similar computational challenges as those faced during the design of the original controller.
Prior approaches have utilized transfer learning to transfer safety guarantees in the form of a barrier certificate while maintaining the control invariant. 
Unfortunately, in practical settings, the source and the target environments often deviate substantially in their control inputs, rendering the aforementioned approach impractical. 
To address this challenge, we propose integrating \emph{inverse dynamics}---a neural network that suggests required action given a desired successor state---of the target system with the barrier certificate of the source system to provide formal proof of safety.
In addition, we propose a validity condition that, when met, guarantees correctness of the controller. 
We demonstrate the effectiveness of our approach through three case studies.

\end{abstract}


\end{frontmatter}

\section{Introduction}
The ever-increasing presence of autonomy in our safety-critical infrastructure---such as self-driving cars, robotics, implantable medical devices, and power grids---has underscored the importance of guaranteed safety in cyber-physical systems.
While formal verification of cyber-physical systems against safety requirements is an undecidable problem~\citep{alur1996hybrid}, deductive verification approaches have shown considerable promise. 
Control barrier certificates (CBCs)~\citep{prajna} are leading deductive approach for effectively synthesizing safe controllers.
CBCs, real-valued functions of the state space, act as a \textit{barrier} between over-approximation of reachable set by a system and unsafe set, guaranteeing safety. 
On the other hand, computing a CBC, if one exists, requires a search in the space of templates of certain form, using optimization methods such as sum-of-squares (SOS)~\citep{parrilo2003semidefinite} optimizations or satisfiability modulo theory (SMT) solvers~\citep{de2011satisfiability}. 
Such search is computationally expensive, and requires a nontrivial understanding of the system and optimization approaches. 
Moreover, systems undergoing even minor changes may render CBCs unusable and necessitate a fresh synthesis of CBCs.
This paper proposes a \emph{transfer learning approach} to transfer a control barrier certificate between systems ``close'' to one-another.

\noindent\textbf{The need for transferring control.}
Existing controllers may require modification for a variety of reasons including: 1) mechanical wear and tear; 2) upgrade in sensors or actuators; 3) different operating conditions such as ambient temperature or pressure; and 4) mismatch between simulation and reality. 
Notice that such changes, while causing a slight discrepancy in control and in guarantees, do not alter the behavior of a system drastically.
Transfer learning provides a learning-based paradigm to adapt to changes precisely in such settings.

\noindent\textbf{Transfer Learning for Control.}
Transfer learning approaches~\citep{weiss2016survey} are concerned with utilizing a previously learned ``knowledge'' in the \emph{source} domain in order to apply it to the \emph{target} domain. 
Classical research in transfer learning was concerned with leveraging learned weights of a neural network from a source domain to speed up training in a related target domain~\citep{bozinovski2020reminder, torrey2010transfer}. 
\cite{christiano2016transfer} proposed an approach to transfer controller from the source domain to the target domain by learning the \emph{inverse dynamics}~\citep{lane1988flight} of the target system as a neural network.
While showing practical success, this approach did not focus on transferring safety guarantees from one domain to another.

\begin{figure*}[t!]
\begin{center}
\includegraphics[width=16cm]{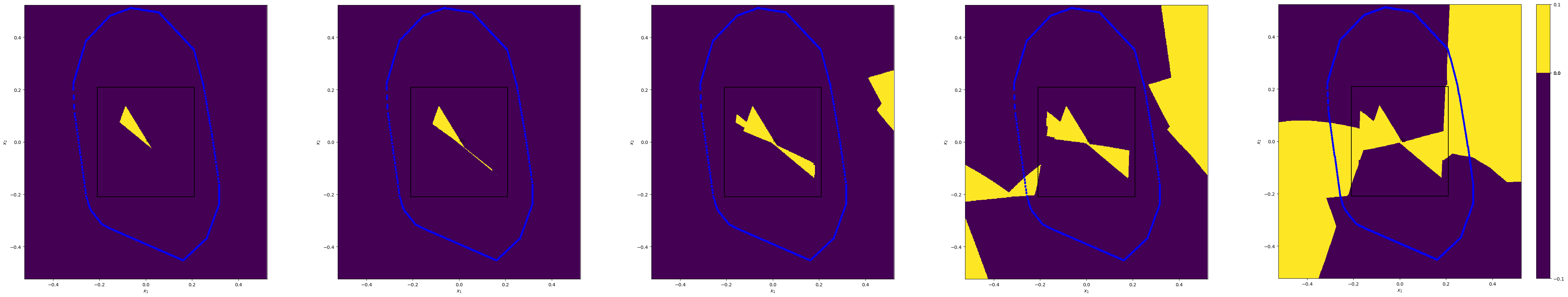}   
\caption{
This figure depicts the dynamics for a series of inverted pendula (angular position on $x$ axis and angular velocity on $y$) differing in their weights and length. 
The leftmost figure is for the source system for which a CBC has already been computed.
We show the gradual failure of the source CBC with changes in the dynamics of an inverted pendulum (details in Section~\ref{sec:expts}).
The blue enclosed region and black square region indicate the zero level set of the CBC and the initial set, respectively.
Moreover, the yellow regions show the violation in a key condition (i.e., decrement in barrier value along transitions) of CBC , while the purple regions show its satisfaction.
} 
\label{fig_failure}
\end{center}
\end{figure*}
\noindent\textbf{Transfer Learning for Safety Guarantees.}
In a recent work,~\citet{tfme} advocated the use of transfer learning to transfer the proof of safety in the form of barrier certificates while using the same controller. 
The application of their approach, while effective in lifting guarantees, is limited to the settings where the source and the target domains are close enough to permit similar control.
We posit that in several practical settings (depicted in Figure~\ref{fig_failure}), two systems may be far enough so as to preclude the use of same control, still may share the ``logical control structure'' to merit a transfer learning of safety guarantees through learning a controller.

\noindent\textbf{Transfer Learning for Control and Safety Guarantees.}
We utilize the idea of learning inverse dynamics~\citep{christiano2016transfer} with the idea of transferring guarantees~\citep{tfme}, to provide an approach to transfer controller while lifting formal guarantees.
Na\"ively, one can learn inverse dynamics first, and then use the approach of~\cite{tfme} to transfer the barrier certificate.
Unfortunately, if for a given inverse dynamics, a barrier certificate cannot be transferred, it provides no further information to modify the inverse dynamics.
To overcome this challenge, we propose barrier certificate guided inverse dynamics algorithm, which utilizes a previously learned CBC for the source system to learn a controller with safety guarantees. 
In particular, we transform the conditions of CBCs into gradients to train an inverse dynamics controller that provides safety guarantees.

\textbf{Contribution.} We proposes a data-driven approach to synthesize provably correct controllers for target systems while taking advantage of previously learned CBCs for source systems. Our approach relies on a learned neural network known as an \textit{inverse dynamics model} to act as a controller for a target system. Moreover, we implement a validity condition within the training of neural networks which ensures correctness of the transferred controller, without the need of post-facto verification. This so-called validity condition is based on Lipschitz continuity of the source and the target dynamics, and the CBC of the source system.
Our method provides a formal proof of safety for a transferred controller, \textit{i.e.}, the closed-loop trajectories of system do not enter the unsafe set.
We illustrate the effectiveness of our algorithm with three case studies.

\textbf{Related Work.} Results in~\citep{prajna2007framework, huang2017probabilistic} expand on CBCs for the safety verification of stochastic systems. Control barrier certificates were then proposed for controller synthesis of deterministic systems~\citep{wieland2007constructive,ames2019control} as well as stochastic ones~\citep{jagtap2020formal,clark2021control}.

There are some drawbacks to the aforementioned methods, as they require mathematical model of a given system. A model of a system is not always available, due to security concerns (one is protecting its intellectual property) or the complexity of the system. 
Moreover, these methods are computationally expensive, since one requires to fix the template of barrier certificates and its controller beforehand (typically in the form of polynomial functions of a certain degree), and then search for its parameters (coefficients of those functions). In most cases, this search relies on optimization methods such as sum-of-squares (SOS)~\citep{parrilo2003semidefinite} optimizations or satisfiability modulo theory (SMT) solvers~\citep{de2011satisfiability}. One often fails to find a barrier certificate with its corresponding controller due to the fix template or computational complexity.
In order to guarantee safety of systems with unknown models, one should rely on data-driven methods. The results in~\citep{nejati2023formal} propose a method based on scenario convex program for safety verification of unknown continuous systems, whereas results in~\citep{nejati2022data} address the controller synthesis.

Neural network-based safety guarantees have gained considerable attention in recent years~\citep{dawson2023safe, zhou2022neural}. Neural networks are universal approximators~\citep{uat} and can represent any Borel-measurable function, therefore they do not suffer from the limitations of a fixed template. Moreover, neural networks are trained via finitely many data points and their training is completely data-driven, thus it does not need a precise mathematical model. One major drawback of parameterizing CBCs and their controllers as neural networks is that it lacks formal guarantee of correctness. In conjunction with finitely many data points for training, one cannot be sure if the CBC conditions are satisfied for the entire state set. Consequently, formal verification of a neural network is required before employing them in safety-critical applications. 
Neural network-based safety certificates have been developed for nonlinear systems in~\citep{zhao2020synthesizing,peruffo2021automated}, for stochastic systems~\citep{mathiesen2022safety}, and for controller synthesis~\citep{jin2020neural}.

\section{Problem Definition}
\label{sec2}
We denote the set of real and non-negative reals by $\R$ and $\R_{\geq 0}$, respectively. 
We denote the cardinality of the set $A$ by $|A|$; and denote and the set difference and Cartesian product of sets $A$ and $B$ by $A \setminus B$ and $A \times B$, respectively. 
We consider $n$-dimensional Euclidean space $\R^n$ equipped with infinity norm, defined as $\| x-y \| = \max_{1 \leq i\leq n}{|x_i{-}y_i|}$ for $x{ =} (x_1,x_2,\ldots,x_n), y {=} (y_1,y_2,\ldots,y_n)\in \R^n$. 
Similarly, we denote Euclidean norm as $\| x-y \|_2 = \sqrt{\sum_{i=1}^n(x_i-y_i)^2}$. 

\subsection{Discrete-time Control Systems}
In this paper, we focus on the safety problem for  discrete-time control systems (dtCS).

\begin{defn}
A discrete-time control system (dtCS) is a tuple $\Sys=(\Xx,U,f)$, where $\Xx \subset \R^n$ represents the state set, $U \subset \R^m $ is the set of inputs, and $f:\Xx \times U \to \Xx$ is the state transition function. 
The evolution of the system under input sequence $u = \seq{u(1),u(2),\ldots}$ is described by:
\begin{equation}
    \Sys: x(t+1) = f(x(t),u(t)),\;\;\; \text{for all }t\in \N.
\end{equation}
We assume the state and input sets, $\Xx, U$, respectively, to be compact, and the map $f$ is unknown but can be simulated via a black-box representation. Moreover, we assume that $f$ is Lipschitz continuous as in the following assumption:
\begin{assum}[Lipschitz Continuity]

Consider a discrete time control system $\Sys = (\Xx,U,f)$. 
The map $f$ is Lipschitz continuous such that for all $x,x^\prime \in \Xx$, and $u,u^\prime\in U$ one gets:
    \begin{equation}
        \| f(x,u)-f(x^\prime,u^\prime)\| \leq \mathcal{L}_x \| x-x^\prime\| + \mathcal{L}_u \|u-u^\prime \|,
    \end{equation}
    for some positive constants $\mathcal{L}_u$ and $\mathcal{L}_x$.
\end{assum}
\end{defn}

\subsection{Safety and Control Barrier Certificate}

A dtCS $\Sys = (\Xx,U,f)$ is safe with respect to initial set of states $\Xx_0 \subseteq \Xx$ and unsafe set $\Xx_u \subseteq \Xx$ if there exists a feedback controller $k:\Xx \rightarrow U$ such that for every trace $\seq{x(0), x(1), \ldots}$, where $x(t+1) = f(x(t),k(x(t))$, we have that $ x(t) \not \in \Xx_u$ for all $t \in \N$. 
We employ the following notion of control barrier certificates (CBCs)~\citep{prajna} which provides sufficient conditions for ensuring safety. 
\begin{defn}
Consider a system $\Sys = (\Xx,U,f)$. A function $B:\Xx\rightarrow \R$ is called a control barrier certificate (CBC) for $\Sys$ with respect to initial set of states $\Xx_0 \subseteq \Xx$ and unsafe set $\Xx_u \subseteq \Xx$ if there exists a controller $k:\Xx {\rightarrow} U$ such that, for some $\eta \in \R_{\geq 0}$, we have: 
\begin{eqnarray}
    B(x) &\leq& -\eta, \;\;\; \forall x \in \Xx_0;\label{eq_barr_1}\\
    B(x)  &>& \eta, \;\;\; \forall x \in \Xx_u; \text{ and } \label{eq_barr_2} \\
     B(f(x,k(x)))-B(x) &\leq& -\eta, \;\;\; \forall x \in \Xx. \label{eq_barr_3}
\end{eqnarray}
The existence of a barrier certificate for $\Sys$ implies that every state sequences starting from $\Xx_0$ under inputs provided by $k$, will never reach $\Xx_u$~\citep{prajna}.
\end{defn}
\subsection{Neural Networks}
Neural networks are universal approximators \citep{uat}. Therefore, these networks can learn any Borel-measurable function based on input-output data. Consider a network with $k$ fully-connected layers where each layer $i$ is characterized with a weight matrix $W_i$ and a bias vector $b_i$ of appropriate size and is followed by an activation function. One can train neural networks on finitely many data points. 

A neural network with $k \in \N$ layers can be viewed as a function $F: \R^{n_i} \to \R^{n_o}$.
Given an input $y_0 \in \R^{n_i}$, a neural network will compute an output $y_k \in \R^{n_o}$ as follows:
\begin{eqnarray*}
y_{1} &=& \sigma (W_1 y_0 + b_1),\\
y_{2} &=& \sigma(W_2 y_1 + b_2),\\
&\vdots &~~~~ \\
y_{k} &=& \sigma (W_k y_{k-1} + b_k).
\end{eqnarray*}
We call $y_{i-1}$ and $y_i$ for $i{\in}\{1,{\ldots},k\}$ the input and output of the $i$-th layer, respectively.
One observes that neural networks with ReLU ($\sigma(x){=}\max(0,x)$) activations describe Lipschitz continuous functions. Moreover, one can obtain the trivial upper bound of Lipschitz constant of a neural network with ReLU activations by multiplying the largest weight of each layer. 
Neural networks are trained on a appropriately defined loss function using a gradient based optimization method~\citep{Goodfellow-et-al-2016}.

\subsection{Problem Definition}

We seek to train a neural network to act as a controller for the target system, while utilizing the CBC and its controller for the source system. The main problem we aim to solve in this paper is formalized below.

\vspace{0.2em}\noindent\textbf{Controller Transfer.} Consider two systems: the \emph{source} $\Sys = (\Xx, U, f)$  and the \emph{target} $\Tt = (\Xx, \hat{U}, \hf)$. Assume that a control barrier certificate $B: \Xx \to \R$ and its corresponding controller $k: \Xx\to U$ with respect to initial and unsafe sets, $\Xx_0, \Xx_u$, respectively, are available for the source system. Furthermore, both CBC and its controller are Lipschitz continuous. 
The \emph{controller transfer problem} is to synthesize a controller $\hat{k}: \Xx \to \hat{U}$ for $\Tt$ to ensure safety of $\Tt$ under $\hat{k}$, with respect to initial and unsafe sets, $\Xx_0,\Xx_u$, respectively
\label{problem_def}
\section{Controller Synthesis using Inverse Dynamics}


\begin{figure*}[h!]
    \centering
    \captionsetup{justification=centering,margin=2cm}
    \resizebox{0.8\textwidth}{!}{
    \begin{tikzpicture}[block/.style = {rectangle, minimum size = 100}, 
    block2/.style = {rectangle, thick, draw, fill=black!5!white, minimum width =100, minimum height = 160},
    block3/.style = {draw, thick, fill=black!5!white, diamond, aspect=2, inner sep=2pt},
    ]
    \tikzstyle{neuron}=[circle,fill=white!80!yellow,draw,minimum size=15pt,inner sep=0pt]      
    \node[block ] (Syss) at (0,-0.75) {\includegraphics[width =0.15\textwidth]{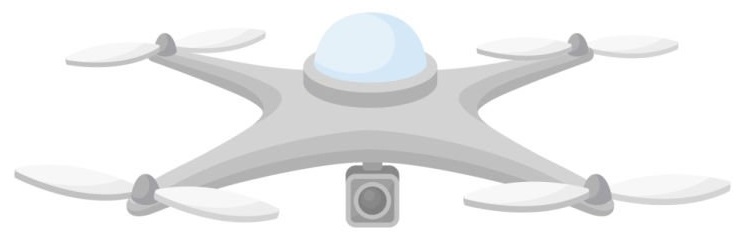}} ;
       
        \node[block] (Spec) at (0,0) {} ;
        \node[block ] (Sys) at (0,-5) {\includegraphics[width =0.15\textwidth]{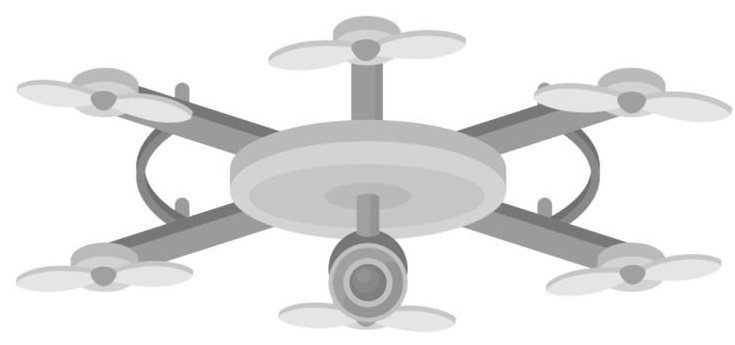}} ;
        \node[block2] (NNBlock) at (6,-2.85) {} ;
    \begin{scope}
            \foreach \y in {1,...,4}
            \node[neuron] (I-\y) at (4.75,-1.2*\y) {};
            \foreach \y in {1,..., 5}
                \node[neuron] (H-\y) at (6, -\y) {};
            \foreach \y in {1,..., 2}
                \node[neuron] (O-\y) at (7.25, -2*\y) {};
        \foreach \i in {1,...,4}
        \foreach \j in {1,...,5}
            \path (I-\i) edge (H-\j);

        \foreach \i in {1,...,5}
        \foreach \j in {1,...,2}
            \path (H-\i) edge (O-\j);
    \end{scope}
    \node[] (CNN) at (6,-6) {\textsf{Inverse Dynamics Controller}}; 

    \node[block3] (LossCheck) at (10,0) {$ {\scriptscriptstyle\mathcal{L}_B\big ( \mathcal{L}_\dagger \frac{\epsilon}{2}  + \mathcal{E} \big )\leq\eta? }$ } ;
    \node[align = center] (6) at (0,-6) {\textsf{Target System} }; 
    \node[align = center] (6) at (0,-1.5) {\textsf{Source System} }; 
    \node[align = center] (Valid) at (15,0) {\textsf{Formal Guarantee} \\ \textsf{of Safety}};
    \node[align = center] (Validity) at (12.5,0.4) {$\mathtt{Yes}$};
    \draw[->] (Sys) -- (NNBlock);
    \draw[->] (7.75,-3) -- (11,-3) -- (11,-7.5)  -- (-3,-7.5)-- (-3,-5) -- (Sys);
    \draw[->] (Spec) -- (NNBlock);
    \draw[->] (NNBlock) -- (LossCheck);
    \draw [->, ultra thick] (LossCheck) -- (Valid);
    \draw[->] (-3,-0.75) -- (Syss);
    \draw[->] (LossCheck) -- (10,1) -- (6,1) --(NNBlock);
    \node[] (Trainagain) at (8, 1.2) {$\mathtt{No}$};
    \node[align = center] (Controlt) at (-2.5,-4.7) {$\hat{k}(x)$};
    \node[align = center] (Controls) at (-2.5,-0.5) {$k(x)$};
    \node[align = center] (nextsteps) at (3,-0.8) {$f(x,k(x))$};
    \node[align = center] (nextstept) at (3,-4.4) {$\hat{f}(x,\hat{k}(x))$};
    \end{tikzpicture}}
     \centering
        \caption{Transfer of Safety Controllers Through Learning Deep Inverse Dynamics Model.}
    
    \label{nnfig}
\end{figure*}
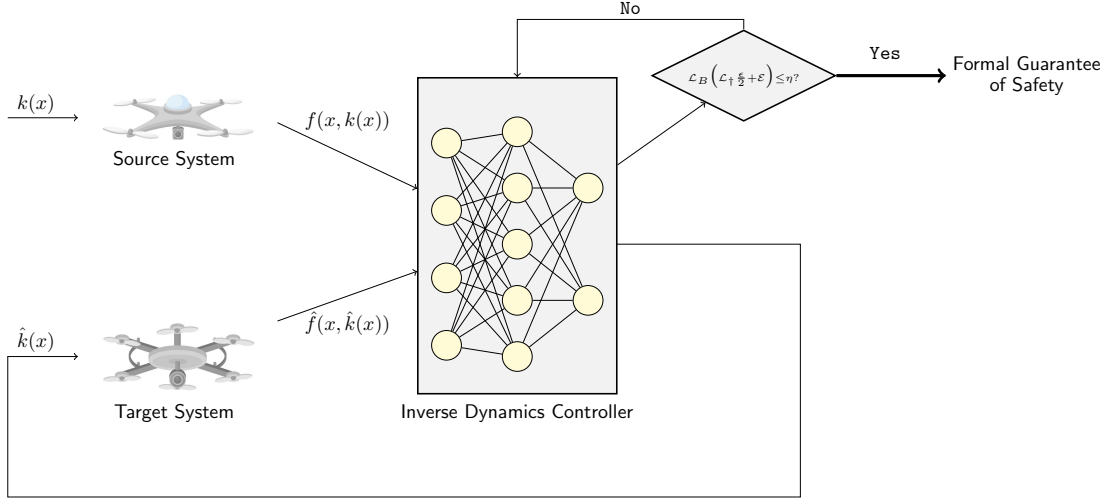

In this section, we propose an algorithm to leverage previously learned CBCs and their corresponding controllers for the source systems, to synthesize a provably correct controller for the target systems which ensures safety.

Since both source and target systems share the same state set, it can be readily observed that conditions~(\ref{eq_barr_1}) and~(\ref{eq_barr_2}) are satisfied for the target system, with the same CBC as in the source system.

Instead of directly applying the controller of the source system to the target one, we aim to transfer high-level properties of the source system's controller and learn the low-level properties using a deep inverse dynamics model.
At each time step, our approach computes the expected actions of the source system's controller.
Instead of implementing these actions on the target system, we simulate the expected next state of the source system. We then rely on a deep inverse dynamics model to determine a suitable input for the target system.
This appropriate action steers the target system toward the expected state of the source system.
Consider the current state of the source system to be $x(t)$ with its corresponding control input $k(x(t))$.
Using the black-box representation of the source system, one can obtain the next state of the source system as $x(t+1) = f(x(t), k(x(t)))$.
Given the CBC for the source system, $x(t+1)$ satisfies condition~(\ref{eq_barr_3}).
Our objective is to learn a controller for the target system such that the next state of the target system matches that of the source system.
In other words, we aim to learn a controller (a.k.a inverse dynamics controller) $\hat{k}(x):\Xx{\rightarrow} \hat{U}$ so that the target system follows the trajectory of the source system at each step, assuming both start from the same initial state.
Since the source system is safe, the target system will never go into unsafe set as well.

In order to generate finitely many training data points, 
we partition the state set $\Xx$ into finitely many cells $\Xx_1,\Xx_2,\ldots,\Xx_M$, by picking a discretization parameter $\epsilon > 0$. We then pick sample points $x_i\in \Xx_i$ from each of these cells such that:
\begin{align}  
    \|x-x_i\| \leq \frac{\epsilon}{2}, \; \text{for all } x \in \Xx_i.
    \label{cover}
\end{align}
Let us denote the set of all those sampled points by $\Xx_d$.
One way of partitioning the state set into such cells, is to partition it into hyperrectangles.
We then pick the centers of these hyperrectangles as representative points.
Finally, we employ the mean squared error (MSE) loss to train the neural network:
\begin{align}
L = \frac{1}{2|\Xx_d|}\sum_{x_i \in \Xx_d} \left(\|f(x_i,k(x_i))-\hat{f}(x_i,\hat{k}(x_i))\|_2^2\right).
\label{loss_function}
\end{align}
Here, $\hat{k}(x_i)$ represents the output of the neural network. Figure~\ref{nnfig} depicts an overview of our method. 

\subsection{Validity Condition}
Neural networks are trained on finitely many data points. Thus, we do no have guarantee for unseen data. 
In order to transfer formal guarantee, we propose a validity condition based on the mismatch between $f$ and $\hat{f}$ over the finitely many data points:
\begin{align}
    \mathcal{E} := \max_{x_i\in \Xx_d} \|f(x_i,k(x_i))-\hat{f}(x_i,\hat{k}(x_i))\|.
    \label{bige}
\end{align}

Now, we state the main theoretical result of the paper that provides a validity condition, under which the trained controller is formally correct.
\vspace{0.2cm}
\begin{thm}
    Consider a source system $\Sys$ = ($\Xx, U,f)$ with a control barrier certificate $B: \Xx \rightarrow \R$ with the corresponding parameter $\eta$ in~(\ref{eq_barr_1})-(\ref{eq_barr_3}), corresponding controller $k: \Xx \rightarrow U$, and a target system $\hat{\Sys}$ = ($\Xx, \hat{U},\hat{f})$. Both $k$ and $B$ are assumed to be Lispchitz continuous with $\mathcal{L}_k$ and $\mathcal{L}_B$ as their Lipschitz constant, respectively. Moreover, we assume that $f$ and $\hat{f}$ satisfy Assumption 2 with Lipschitz constants $\mathcal{L}_x, \mathcal{L}_u$ for the source system and $\mathcal{L}_{\hat{x}}, \mathcal{L}_{\hat{u}}$ for the target one, respectively. Let $\Xx_d$ be a finite set of sampled data points according to~(\ref{cover}) with discretization parameter $\epsilon$, and $\Xx_0,\Xx_u \subseteq \Xx$ to be the corresponding set of initial and unsafe states, respectively. The target system with an inverse dynamics controller $\hat{k}:\Xx \rightarrow \hat{U}$ synthesized according to Algorithm 1, is guaranteed to be safe.
    
\end{thm}
\begin{pf}
As mentioned in~(\ref{cover}), for every $x \in \Xx$, there exists $x_i \in \Xx_d$ such that $\|x-x_i\|\leq \frac{\epsilon}{2}$. Thus, one obtains:
\begin{align}
    \nonumber &\|f(x,k(x)))-\hat{f}(x,\hat{k}(x))\| \;  \\[5pt]
    \nonumber  \leq& \|f(x,k(x))-f(x_i,k(x_i)) + f(x_i,k(x_i)) \\[5pt] 
    & -\hat{f}(x,\hat{k}(x)) + \hat{f}(x_i,k(x_i)) - \hat{f}(x_i,\hat{k}(x_i))\| \label{eq9th}  \\[5pt]
    \nonumber \leq& \|f(x,k(x))-f(x_i,k(x_i)) \| 
    + \| \hat{f}(x_i,\hat{k}(x_i))-\hat{f}(x,\hat{k}(x))\| \\[5pt]
    &  + \| f(x_i,k(x_i)) - \hat{f}(x_i,\hat{k}(x_i))\| \label{eq10th}  \\[5pt]
    \nonumber  \leq & \mathcal{L}_x \|x{-}x_i\| + \mathcal{L}_{\hat{x}} \|x{-}x_i\| +\mathcal{L}_u\|k(x){-}k(x_i) \| \\[5pt]
    &+\mathcal{L}_{\hat{u}}\| \hat{k}(x) {-} \hat{k}(x_i) \| + \|f(x_i,k(x_i)){-}\hat{f}(x_i,\hat{k}(x_i))\|,\label{eq11th}
\end{align}
for all $ x\in \Xx,\; x_i\in \Xx_d$. 
Here, the inequality (\ref{eq10th}) follows from the triangular inequality and the inequality (\ref{eq11th}) follows from Lipschitz continuity. 
Since both $k(x)$ and $\hat{k}(x)$ are Lipschitz continuous with some constants $\mathcal{L}_k$ and $\mathcal{L}_{\hat{k}}$, respectively, we have:
\begin{align}
    &\nonumber \mathcal{L}_u\|k(x)-k(x_i) \|  +\mathcal{L}_{\hat{u}}\| \hat{k}(x) - \hat{k}(x_i) \|  \leq \\[5pt]
    &\mathcal{L}_u \mathcal{L}_k \|x-x_i\|  +\mathcal{L}_{\hat{u}} \mathcal{L}_{\hat{k}} \|x-x_i\|, \;\;\; \text{for all } x\in \Xx, x_i \in \Xx_d.
\end{align}
Thus, according to~(\ref{eq11th}), one obtains:
\[
\|f(x,k(x))){-}\hat{f}(x,\hat{k}(x))\| \leq \left( \mathcal{L}_x {+} \mathcal{L}_u\mathcal{L}_k {+} \mathcal{L}_{\hat{x}} {+} \mathcal{L}_{\hat{u}}\mathcal{L}_{\hat{k}} \right) \frac{\epsilon}{2}  {+} \mathcal{E}, 
\]    
for all $ x\in \Xx,\; x_i\in \Xx_d$, 
$\| x{-}x_i  \| {\leq} \frac{\epsilon}{2}$, and $\mathcal{E}$ is defined as~(\ref{bige}). 

We employ the CBC of the source system for the target one. It can be readily verified that conditions~(\ref{eq_barr_1}) and~(\ref{eq_barr_2}) hold for the target system. 
Let us focus our attention on condition~(\ref{eq_barr_3}):
\begin{align}
    &\nonumber B(\hat{f}(x,\hat{k}(x)))-B(x)\\
    \leq&\nonumber  B(\hat{f}(x,\hat{k}(x))) {-}B(f(x,k(x))) {+} B(f(x,k(x))){-}B(x) \\ 
     \leq&\nonumber B(\hat{f}(x,\hat{k}(x)))  {-} B(f(x,k(x))) {-} \eta \\[5pt]
     \leq& \nonumber\mathcal{L}_B \|f(x,k(x)) - \hat{f}(x,\hat{k}(x))\| -\eta \\[5pt]
    \leq & \mathcal{L}_B\left( ( \mathcal{L}_x + \mathcal{L}_u\mathcal{L}_k  +  \mathcal{L}_{\hat{x}} + \mathcal{L}_{\hat{u}}\mathcal{L}_{\hat{k}}  ) \frac{\epsilon}{2}  + \mathcal{E} \right) -\eta, \label{barr_target_three}
\end{align}
for all $x\in \Xx$, where $\mathcal{L}_B$ is the Lipschitz constant of the CBC, and $\eta$ is its corresponding parameter in~(\ref{eq_barr_1})-(\ref{eq_barr_3}). In order to satisfy condition~(\ref{eq_barr_3}) for the target system, inequality~(\ref{barr_target_three}) must be less or equal to $0$. Thus, one needs:
\begin{align}
   &\mathcal{L}_B\left (\mathcal{L}_{\dagger} \frac{\epsilon}{2}  + \mathcal{E} \right ) -\eta \leq 0,\label{dagger}
 \end{align}
where $\mathcal{L}_{\dagger} {=} \mathcal{L}_x + \mathcal{L}_u\mathcal{L}_k  + \mathcal{L}_{\hat{x}} + \mathcal{L}_{\hat{u}}\mathcal{L}_{\hat{k}}$,
which is satisfied based on Algorithm 1. Therefore, $B$ together with controller $\hat{k}$ is a CBC for the target system.\hspace{3.8cm} \qed 
\end{pf}

The pseudo-code of our proposed method is introduced in Algorithm 1. Note that the guarantees provided by our approach are subject to the termination of this algorithm.

\begin{algorithm}
\caption{Learning the inverse dynamics controller}\label{alg:cap}
\textbf{Input:} $\Xx_0, \Xx_u, \Xx, \epsilon, B, \eta, k,\mathcal{L}_x,\mathcal{L}_{\hat{x}},\mathcal{L}_u,\mathcal{L}_{\hat{u}}, \mathcal{L}_k,\mathcal{L}_B$, Neural network architecture  \\
\textbf{Output} $\hat{k}$
\begin{algorithmic}
\State Construct the training set $\Xx_d$.
\State Initialize neural network's ($\hat{k}$) weights and biases.
\State $\mathcal{E} \gets  \max_{x_i\in \Xx_d} \|f(x_i,k(x_i))-\hat{f}(x_i,\hat{k}(x_i))\|$
\State $\mathcal{L}_{\hat{k}} \gets \text{Lipschitz constant of } \hat{k} $
    \State $\mathcal{L}_\dagger \gets \big ( \mathcal{L}_x + \mathcal{L}_u\mathcal{L}_k \big ) + \big ( \mathcal{L}_{\hat{x}} + \mathcal{L}_{\hat{u}}\mathcal{L}_{\hat{k}} \big )$
\While{$ \mathcal{L}_B\Big ( \mathcal{L}_\dagger \frac{\epsilon}{2}  + \mathcal{E} \Big )>\eta$}
    \State $L \gets \frac{1}{2|\Xx_d|}\sum_{x_i \in \Xx_d} \|f(x_i,k(x_i))-\hat{f}(x_i,\hat{k}(x_i))\|^2_2 $
    \State Train the neural network $\hat{k}$ based on $L$.
    \State $\mathcal{E} \gets  \max_{x_i\in \Xx_d} \|f(x_i,k(x_i))-\hat{f}(x_i,\hat{k}(x_i))\|$
    \State $\mathcal{L}_{\hat{k}} \gets \text{Lipschitz constant of } \hat{k} $
    \State $\mathcal{L}_\dagger \gets \big ( \mathcal{L}_x + \mathcal{L}_u\mathcal{L}_k \big ) + \big ( \mathcal{L}_{\hat{x}} + \mathcal{L}_{\hat{u}}\mathcal{L}_{\hat{k}} \big )$
\EndWhile
\State Return $\hat{k}$
\end{algorithmic}
\end{algorithm}

\section{Experiments}
\label{sec:expts}
In this section, we illustrate the effectiveness of our algorithm with three case studies. All experiments are conducted on a Nvidia RTX 4090 GPU coupled with an Intel core i7-13700k CPU, and 32GB of DDR5 RAM. For the inverse dynamics model, we consider a neural network with 4 hidden layers, each containing 200 neurons. The dimensions of the input and output layers depend on $\Xx$ and $\hat{U}$, respectively. Moreover, we use Adam optimizer~\citep{zhang2018improved} to train the network with a learning rate of $5*10^{-6}$.

We assume that Lipschitz constants of the CBC, $f$, and $\hat{f}$ are known.
If the Lipschitz constants are unknown, one can leverage sampling methods such as~\citep{wood1996estimation,strongin2019acceleration,calliess2020lazily} to estimate those constants. Furthermore, a neural network can be forced to have a small Lipschitz constant by adding a regularization term to the loss~\citep{Goodfellow-et-al-2016}.

For all case studies, we adopted the results in~\citep{mahathi} to obtain a CBC and its corresponding controller for source systems. 

The following table summarizes the effectiveness of our algorithm. For a more detailed explanation, we refer readers to each case study's corresponding section.
\begin{table}[h!]
\caption{Computation time for each case study}
\label{table_1}
\begin{center}
\begin{tabularx}{\columnwidth}{*{4}{X}}
\toprule
Case Study & Computation time (minute) & Transfer learning (minute)\\
\midrule
Inverted Pendulum &$120$& $\mathbf{3}$\\
\midrule
DC Motor&$30$&$\mathbf{1.5}$\\
\midrule
Quadrotor Drone&$360$&$\mathbf{2}$\\
\bottomrule
\end{tabularx}
\end{center}
\end{table}

It is important to emphasize that in our case studies, when referring to the model of the source system, we are treating it solely as a black-box representation for the simulation purposes. We did not incorporate the model to encode the conditions of control barrier certificates.

\subsection{Inverted Pendulum}
\begin{figure*}[h!]
\centering
\begin{subfigure}{0.42 \textwidth}
\includegraphics[width=7.5cm]{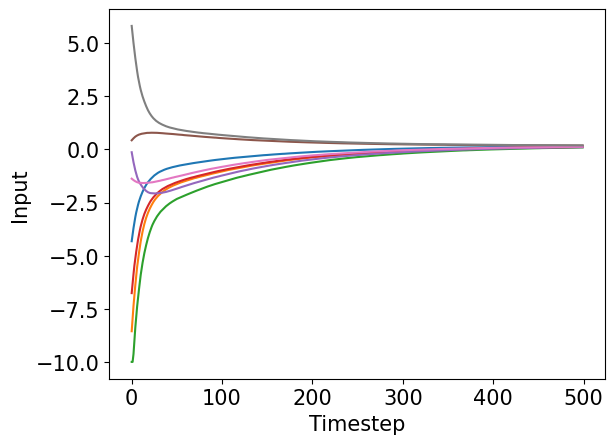}
\caption{}
\label{fig_inp}
\end{subfigure}
\centering
\begin{subfigure}{0.42\textwidth}
\includegraphics[width=7.5cm]{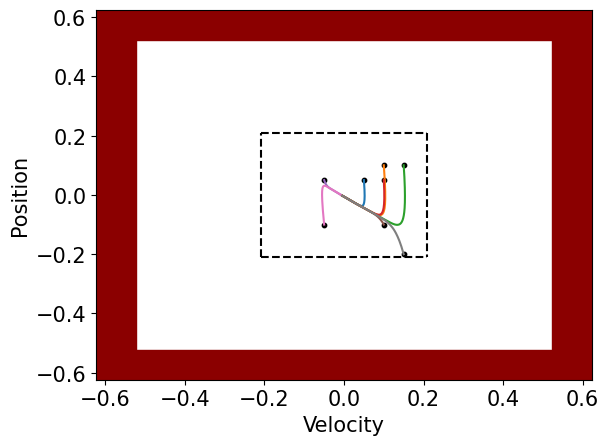}    
\caption{}
\label{fig_traj}
\end{subfigure}
\caption{Some state sequences showing the evolution of the target inverted pendulum (Figure \ref{fig_traj}) and the corresponding input (Figure \ref{fig_inp}).
The areas marked with red indicate the unsafe set. We denote the initial set by the dotted black square. }
\label{}
\end{figure*}

We consider the source system $\Sys=(\Xx,U,f)$ to be an inverted pendulum where $\Xx=[\frac{-\pi}{4},\frac{\pi}{4}]\times[\frac{-\pi}{4},\frac{\pi}{4}]$, $\Xx_0=[\frac{-\pi}{15},\frac{\pi}{15}]\times[\frac{-\pi}{15},\frac{\pi}{15}]$, and $\Xx_u = \Xx \setminus [\frac{-\pi}{6},\frac{\pi}{6}]\times[\frac{-\pi}{6},\frac{\pi}{6}]$. The transition function is given by:
\begin{align}
f(x_1,x_2) = \left[\begin{array}{c}
x_1+\tau x_2 \\
x_2+\frac { g\tau } { l }\sin \left(x_1+\frac{1}{m l^2} k(x)\right) 
\end{array}\right], \nonumber
\end{align}
where $x_1$ and $x_2$ are the angular position and velocity, respectively. Moreover, $g = 9.8$ is the gravitational acceleration, and $l = 1$ and $m = 1$ are the length and mass of the pendulum, respectively. Constant $\tau = 0.01$ is the sampling rate, and constants $\mathcal{L}_x=1.1$, $\mathcal{L}_B=2$, and $\eta = 0.07637$ are Lipschitz constants according to~(\ref{dagger}), Lipschitz constant of $B$, and its corresponding parameter, respectively. For the target system, we choose $l = 1.5$ and $m=1.5$. After training the neural network, we get $\mathcal{L}_{\dagger} = 2.2$. For both systems, the discretization parameter and input set are $\epsilon = 9*10^{-4}$, and $U=\hat{U}=[-10,10]$, respectively. Some state sequences and their corresponding inputs are depicted in Figure~\ref{fig_traj} and Figure~\ref{fig_inp}, respectively.

\begin{figure*}[ht!]
\begin{center}
\includegraphics[width=16cm]{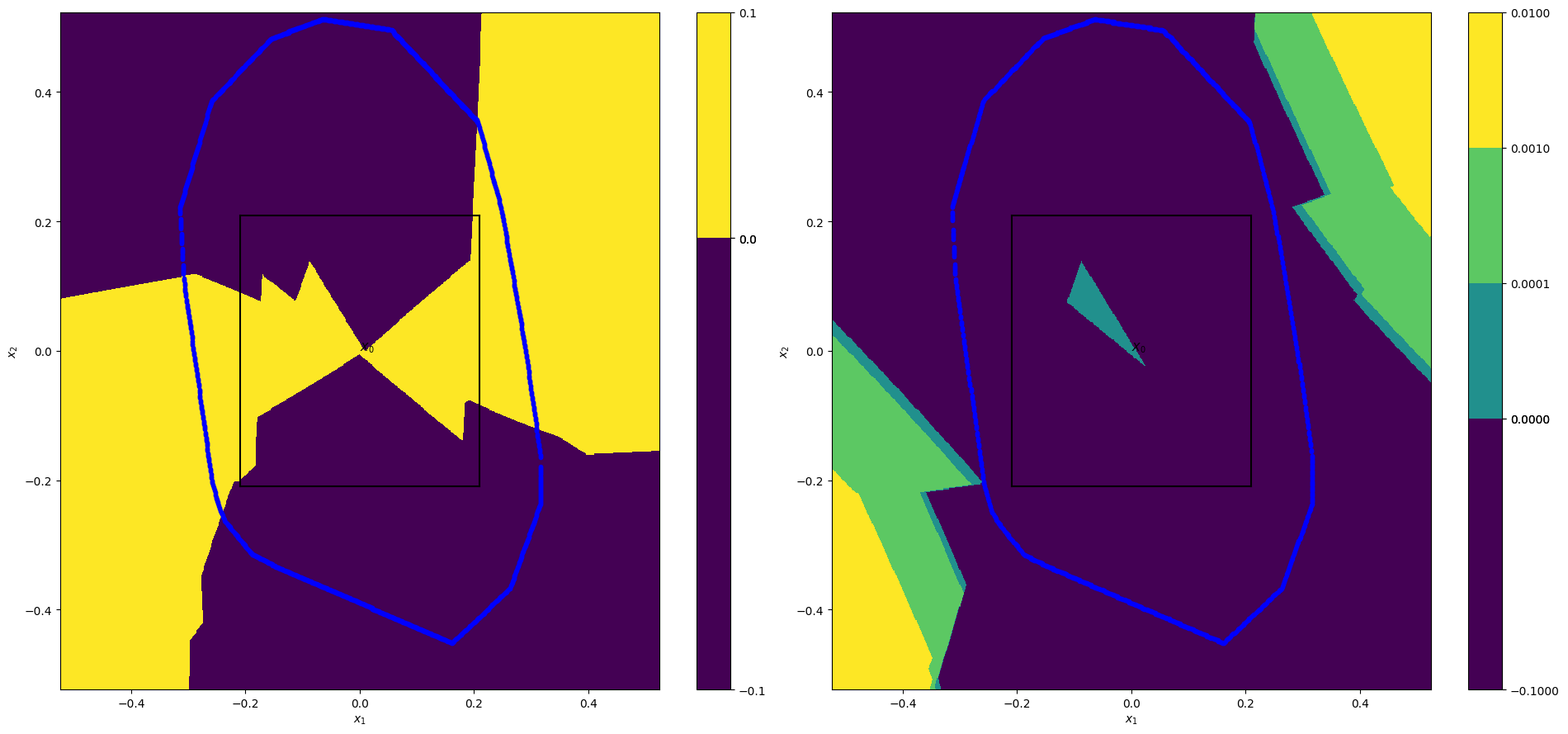}    
\caption{Barrier certificate for the target system.
Figure on the left indicates the value of the CBC when using the control policy of the source system, whereas Figure on the right depicts the same CBC but with the learned inverse dynamics controller.} 
\label{fig1}
\end{center}
\end{figure*}

In this experiment, if one uses the same controller, the CBC does not provide a guarantee of safety, and based on simulations, the target system enters the unsafe set. Our training method converged with 10000 iterations in 3 minutes, with $\mathcal{E} = 2.5*10^{-4}$. Note that computing a control barrier certificate with its corresponding controller from scratch, takes roughly about 1.5 to 2 hours.
The CBC values for both controllers are depicted in Figure~\ref{fig1}.

\subsection{DC Motor}

In this case study, we consider a discrete-time DC motor $\Sys =( \Xx,U,f)$, with the transition function:

\begin{align}
f(x_1,x_2) = \left[\begin{array}{c}
x_1+\tau\left(\frac{-R}{L} x_1-\frac{K}{L} x_2+\frac{1}{L} k(x)\right) \\
x_2+\tau\left(\frac{K}{J} x_1-\frac{b}{J} x_2\right)
\end{array}\right],
\end{align}
where $x_1$ and $x_2$ are the armature current and rotational speed of the shaft, respectively. The parameters of the source system are $R=1,\; L=0.5,\; J=0.05$, and $b=1$, which represent the electric resistance, the electric inductance, the moment of inertia of the rotor, the friction constant, respectively. Moreover, $K=0.01$ denotes both the motor torque and electromotive force constant. Here, $\tau = 0.01$ is the sampling time. The regions of interest are, $\Xx=[-0.7,0.7]\times[-0.1,0.1]$, $\Xx_0 = [-0.005, 0.005]\times[-0.05,0.05]$, and $\Xx_u = [0.5,0.7]\times[0.06,1]$, respectively. The input voltage $u$ for both systems is bounded within $U=\hat{U}=[-1,1]$. Furthermore, $\mathcal{L}_x=1$ and $\mathcal{L}_u=0.02$ are the Lipschitz constants of the source system. For the target system, we consider $L=0.55$ and $R=1.2$, and the remaining parameters the same as the source system, and $\epsilon = 0.0004$ as the discritization parameter. The Lipschitz constant of the CBC and its parameter are $\mathcal{L}_B = 10$, and $\eta = 0.0211$, respectively. Here, we get $\mathcal{L}_\dagger = 2.2$, and our algorithm converged with $10000$ iterations and $\mathcal{E} = 10^{-3}$ in 1.5 minutes. Training a new CBC with its controller for the target system takes roughly around 30-45 minutes. 
\subsection{Quadrotor Drone}
\begin{figure*}[htb!]
\centering
\begin{subfigure}{0.42 \textwidth}
\includegraphics[width=7.5cm]{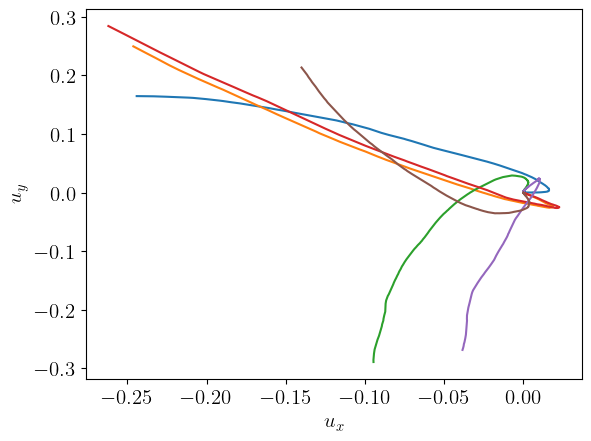}
\caption{}
\label{fig_inp_drone}
\end{subfigure}
\centering
\begin{subfigure}{0.42\textwidth}
\includegraphics[width=7.5cm]{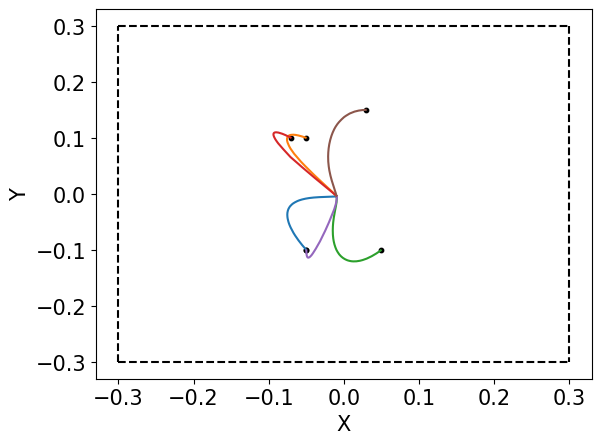}    
\caption{}
\label{fig_traj_drone}
\end{subfigure}
\caption{Some state sequences showing the evolution of the target drone (Figure \ref{fig_traj}) and the corresponding input (Figure \ref{fig_inp}).
We denote the initial set by the dotted black square. }
\label{}
\end{figure*}

For the last case study, we consider a 4 dimensional drone, burrowed from~\citep{zhong2023towards}. The state transition function of the source system is in the form of $f(x,u)= Ax + Bu$, for all $x \in \Xx, u \in U$, with matrices $A,B$ as follows:
\begin{equation}
    A:=\begin{bmatrix}
1 & \tau & 0 &0\\
 0&1  &0  &0 \\
 0&0  &1  &\tau \\
 0&0  &0  &1 
\end{bmatrix}, 
    B:=\begin{bmatrix}
\frac{\tau^2}{2} & 0 \\
 \tau&0   \\
 0&\frac{\tau^2}{2}  \\
 0&\tau   
 \end{bmatrix}, 
\end{equation}
where $\tau =0.01$ is the sampling time. Vectors $x:=[x_x;v_x;x_y;v_y]$, and $u:= [u_x;u_y]$ denote the state and the control input of the drone, respectively, with $x_i$, $v_i$, and $u_i$ being the position, velocity, and acceleration of the quadrotor drone on the $i$ axis, $i\in \{x,y\}$, respectively. Furthermore, $\Xx = [-3,3]^4$, $\Xx_0 =[-0.3,0.3]^4$, $\Xx_u = \Xx \setminus [-2,2]^4$ are state set, initial state set, and unsafe set, respectively. The Lischpitz constant of the CBC and its parameter are $\mathcal{L}_B = 0.269$ and $\eta = 0.1$, respectively. Here, we obtain $\mathcal{L}_{\dagger} = 2.02$, and $\mathcal{E} = 7*10^{-4}$. We discretize the state set with $\epsilon=0.2$, and for both systems $\hat{U} = U = [-2,2]^2$. For the target system, we negated entries of the matrix $B$, to model the differences in the actuation. If one uses the controller for the source system, it will steer the target system to the unsafe set. Some state trajectories of the drone and their corresponding control inputs are depicted Figure~\ref{fig_traj_drone} and Figure~\ref{fig_inp_drone}, respectively. 
Our algorithm converged with 10000 iterations in 2 minutes, as opposed to 6 hours for synthesizing a new CBC for the target system.

\section{Conclusion and Future work}

We proposed a data-driven approach to synthesize provably correct controllers for target systems while leveraging previously learned control barrier certificates for source systems. Our approach relies on a trained neural network, referred to as an "inverse dynamics model," to serve as a controller for a target system. Additionally, we have incorporated a validity condition into the neural network training process, ensuring the correctness of the transferred controller without the necessity for post-facto verification. We demonstrated the effectiveness of our algorithm through three case studies.
For future work, we intend to transfer guarantees in stochastic systems. Another direction is to go beyond safety and transfer controllers for liveness specifications. 

\bibliography{root}            

\begin{thebibliography}{32}
\providecommand{\natexlab}[1]{#1}
\providecommand{\url}[1]{\texttt{#1}}
\providecommand{\urlprefix}{URL }
\expandafter\ifx\csname urlstyle\endcsname\relax
  \providecommand{\doi}[1]{doi:\discretionary{}{}{}#1}\else
  \providecommand{\doi}{doi:\discretionary{}{}{}\begingroup \urlstyle{rm}\Url}\fi

\bibitem[{Alur et~al.(1996)Alur, Henzinger, and Sontag}]{alur1996hybrid}
Alur, R., Henzinger, T.A., and Sontag, E.D. (1996).
\newblock \emph{Hybrid systems III: Verification and control}, volume~3.
\newblock Springer Science \& Business Media.

\bibitem[{Ames et~al.(2019)Ames, Coogan, Egerstedt, Notomista, Sreenath, and Tabuada}]{ames2019control}
Ames, A.D., Coogan, S., Egerstedt, M., Notomista, G., Sreenath, K., and Tabuada, P. (2019).
\newblock Control barrier functions: Theory and applications.
\newblock In \emph{2019 18th European control conference (ECC)}, 3420--3431. IEEE.

\bibitem[{Anand and Zamani(2023)}]{mahathi}
Anand, M. and Zamani, M. (2023).
\newblock {Formally verified neural network control barrier certificates for unknown systems}.
\newblock In \emph{{Proceedings of the 22nd World Congress of the International Federation of Automatic Control}}, 2742--2747.

\bibitem[{Bozinovski(2020)}]{bozinovski2020reminder}
Bozinovski, S. (2020).
\newblock Reminder of the first paper on transfer learning in neural networks, 1976.
\newblock \emph{Informatica}, 44(3).

\bibitem[{Calliess et~al.(2020)Calliess, Roberts, Rasmussen, and Maciejowski}]{calliess2020lazily}
Calliess, J.P., Roberts, S.J., Rasmussen, C.E., and Maciejowski, J. (2020).
\newblock Lazily adapted constant kinky inference for nonparametric regression and model-reference adaptive control.
\newblock \emph{Automatica}, 122, 109216.

\bibitem[{Christiano et~al.(2016)Christiano, Shah, Mordatch, Schneider, Blackwell, Tobin, Abbeel, and Zaremba}]{christiano2016transfer}
Christiano, P., Shah, Z., Mordatch, I., Schneider, J., Blackwell, T., Tobin, J., Abbeel, P., and Zaremba, W. (2016).
\newblock Transfer from simulation to real world through learning deep inverse dynamics model.
\newblock \emph{arXiv preprint arXiv:1610.03518}.

\bibitem[{Clark(2021)}]{clark2021control}
Clark, A. (2021).
\newblock Control barrier functions for stochastic systems.
\newblock \emph{Automatica}, 130, 109688.

\bibitem[{Dawson et~al.(2022)Dawson, Gao, and Fan}]{dawson2023safe}
Dawson, C., Gao, S., and Fan, C. (2022).
\newblock Safe control with learned certificates: A survey of neural {L}yapunov, barrier, and contraction methods for robotics and control.
\newblock \emph{IEEE Transactions on Robotics}, 39, 1749--1767.

\bibitem[{De~Moura and Bj{\o}rner(2011)}]{de2011satisfiability}
De~Moura, L. and Bj{\o}rner, N. (2011).
\newblock Satisfiability modulo theories: introduction and applications.
\newblock \emph{Communications of the ACM}, 54(9), 69--77.

\bibitem[{Goodfellow et~al.(2016)Goodfellow, Bengio, and Courville}]{Goodfellow-et-al-2016}
Goodfellow, I., Bengio, Y., and Courville, A. (2016).
\newblock \emph{Deep Learning}.
\newblock MIT Press.

\bibitem[{Hornik et~al.(1989)Hornik, Stinchcombe, and White}]{uat}
Hornik, K., Stinchcombe, M., and White, H. (1989).
\newblock Multilayer feedforward networks are universal approximators.
\newblock \emph{Neural Networks}, 2(5), 359--366.

\bibitem[{Huang et~al.(2017)Huang, Chen, Lin, Yang, and Li}]{huang2017probabilistic}
Huang, C., Chen, X., Lin, W., Yang, Z., and Li, X. (2017).
\newblock Probabilistic safety verification of stochastic hybrid systems using barrier certificates.
\newblock \emph{ACM Transactions on Embedded Computing Systems (TECS)}, 16(5s), 1--19.

\bibitem[{Jagtap et~al.(2020)Jagtap, Soudjani, and Zamani}]{jagtap2020formal}
Jagtap, P., Soudjani, S., and Zamani, M. (2020).
\newblock Formal synthesis of stochastic systems via control barrier certificates.
\newblock \emph{IEEE Transactions on Automatic Control}, 66(7), 3097--3110.

\bibitem[{Jin et~al.(2020)Jin, Wang, Yang, and Mou}]{jin2020neural}
Jin, W., Wang, Z., Yang, Z., and Mou, S. (2020).
\newblock Neural certificates for safe control policies.
\newblock \emph{arXiv preprint arXiv:2006.08465}.

\bibitem[{Lane and Stengel(1988)}]{lane1988flight}
Lane, S.H. and Stengel, R.F. (1988).
\newblock Flight control design using non-linear inverse dynamics.
\newblock \emph{Automatica}, 24(4), 471--483.

\bibitem[{Mathiesen et~al.(2022)Mathiesen, Calvert, and Laurenti}]{mathiesen2022safety}
Mathiesen, F.B., Calvert, S.C., and Laurenti, L. (2022).
\newblock Safety certification for stochastic systems via neural barrier functions.
\newblock \emph{IEEE Control Systems Letters}, 7, 973--978.

\bibitem[{Nadali et~al.(2023)Nadali, Trivedi, and Zamani}]{tfme}
Nadali, A., Trivedi, A., and Zamani, M. (2023).
\newblock {Transfer Learning for Barrier Certificates}.
\newblock In \emph{{62nd IEEE Conference on Decision and Control}}.

\bibitem[{Nejati et~al.(2023)Nejati, Lavaei, Jagtap, Soudjani, and Zamani}]{nejati2023formal}
Nejati, A., Lavaei, A., Jagtap, P., Soudjani, S., and Zamani, M. (2023).
\newblock Formal verification of unknown discrete- and continuous-time systems: A data-driven approach.
\newblock \emph{IEEE Transactions on Automatic Control}, 68(5), 3011--3024.

\bibitem[{Nejati et~al.(2022)Nejati, Zhong, Caccamo, and Zamani}]{nejati2022data}
Nejati, A., Zhong, B., Caccamo, M., and Zamani, M. (2022).
\newblock Data-driven controller synthesis of unknown nonlinear polynomial systems via control barrier certificates.
\newblock In \emph{Learning for Dynamics and Control Conference}, 763--776. PMLR.

\bibitem[{Parrilo(2003)}]{parrilo2003semidefinite}
Parrilo, P.A. (2003).
\newblock Semidefinite programming relaxations for semialgebraic problems.
\newblock \emph{Mathematical programming}, 96, 293--320.

\bibitem[{Peruffo et~al.(2021)Peruffo, Ahmed, and Abate}]{peruffo2021automated}
Peruffo, A., Ahmed, D., and Abate, A. (2021).
\newblock Automated and formal synthesis of neural barrier certificates for dynamical models.
\newblock In \emph{International conference on tools and algorithms for the construction and analysis of systems}, 370--388. Springer.

\bibitem[{Prajna and Jadbabaie(2004)}]{prajna}
Prajna, S. and Jadbabaie, A. (2004).
\newblock Safety verification of hybrid systems using barrier certificates.
\newblock In \emph{Hybrid Systems: Computation and Control}, 477--492. Springer Berlin Heidelberg, Berlin, Heidelberg.

\bibitem[{Prajna et~al.(2007)Prajna, Jadbabaie, and Pappas}]{prajna2007framework}
Prajna, S., Jadbabaie, A., and Pappas, G.J. (2007).
\newblock A framework for worst-case and stochastic safety verification using barrier certificates.
\newblock \emph{IEEE Transactions on Automatic Control}, 52(8), 1415--1428.

\bibitem[{Strongin et~al.(2019)Strongin, Barkalov, and Bevzuk}]{strongin2019acceleration}
Strongin, R., Barkalov, K., and Bevzuk, S. (2019).
\newblock Acceleration of global search by implementing dual estimates for {L}ipschitz constant.
\newblock In \emph{International Conference on Numerical Computations: Theory and Algorithms}, 478--486. Springer.

\bibitem[{Torrey and Shavlik(2010)}]{torrey2010transfer}
Torrey, L. and Shavlik, J. (2010).
\newblock Transfer learning.
\newblock In \emph{Handbook of research on machine learning applications and trends: algorithms, methods, and techniques}, 242--264. IGI global.

\bibitem[{Weiss et~al.(2016)Weiss, Khoshgoftaar, and Wang}]{weiss2016survey}
Weiss, K., Khoshgoftaar, T.M., and Wang, D. (2016).
\newblock A survey of transfer learning.
\newblock \emph{Journal of Big data}, 3(1), 1--40.

\bibitem[{Wieland and Allg{\"o}wer(2007)}]{wieland2007constructive}
Wieland, P. and Allg{\"o}wer, F. (2007).
\newblock Constructive safety using control barrier functions.
\newblock \emph{IFAC Proceedings Volumes}, 40(12), 462--467.

\bibitem[{Wood and Zhang(1996)}]{wood1996estimation}
Wood, G. and Zhang, B. (1996).
\newblock Estimation of the {L}ipschitz constant of a function.
\newblock \emph{Journal of Global Optimization}, 8, 91--103.

\bibitem[{Zhang(2018)}]{zhang2018improved}
Zhang, Z. (2018).
\newblock Improved {A}dam optimizer for deep neural networks.
\newblock In \emph{2018 IEEE/ACM 26th international symposium on quality of service (IWQoS)}, 1--2.

\bibitem[{Zhao et~al.(2020)Zhao, Zeng, Chen, and Liu}]{zhao2020synthesizing}
Zhao, H., Zeng, X., Chen, T., and Liu, Z. (2020).
\newblock Synthesizing barrier certificates using neural networks.
\newblock In \emph{Proceedings of the 23rd international conference on hybrid systems: computation and control}, 1--11.

\bibitem[{Zhong et~al.(2023)Zhong, Liu, Caccamo, and Zamani}]{zhong2023towards}
Zhong, B., Liu, S., Caccamo, M., and Zamani, M. (2023).
\newblock Towards trustworthy {AI}: Sandboxing {AI}-based unverified controllers for safe and secure cyber-physical systems.
\newblock In \emph{2023 62nd IEEE Conference on Decision and Control (CDC)}, 1833--1840.

\bibitem[{Zhou et~al.(2022)Zhou, Quartz, De~Sterck, and Liu}]{zhou2022neural}
Zhou, R., Quartz, T., De~Sterck, H., and Liu, J. (2022).
\newblock Neural {L}yapunov control of unknown nonlinear systems with stability guarantees.
\newblock \emph{Advances in Neural Information Processing Systems}, 35, 29113--29125.

\end{thebibliography}
                                                   







\end{document}